# HVAF Spraying of NiTi Coatings: Microstructure, Phase Transformation and Shape Memory Behavior

Sneha Samal[1], Shrikant Joshi[2], Stefan Björklund[2], Mohit Chandra[1], Akhil Bhardwaj[1], Jaromír Kopeček[1], Stanislav Habr[1], Lukáš Václavek[3,4], Jan Tomáštík[3,4], Ondřej Tyc[1], and Petr Šittner[1]

[1]Institute of Physics of the Czech Academy of Sciences, Na Slovance 1992/2, Prague, 18200, Czech Republic

[2]Department of Engineering Science, University West, 46186, Trollhättan, Sweden

[3]Palacký University in Olomouc, Faculty of Science, Joint Laboratory of Optics of Palacký University and Institute of Physics AS CR, 17. listopadu 12, 779 00 Olomouc, Czech Republic

[4]Institute of Physics of the Czech Academy of Sciences, Joint Laboratory of Optics of Palacký University and Institute of Physics of the Czech Academy of Sciences, 779 00 Olomouc, Czech Republic

**Abstract**

Depositing various coatings on surface of engineering components with the aim to improve their performance concerning wear, corrosion, friction, and thermal protection is already a standard practice. Depositing metallic NiTi shape memory alloy coatings may be a viable alternative for hard ceramic coatings. NiTi coatings offer additional benefits originating from unique functional thermomechanical properties. However, fabrication of thick NiTi coatings turned out to be difficult. Standard electroplating and laser cladding methods are not suitable for NiTi, the most widely used plasma spray methods tend to produce chemically inhomogeneous coatings that do not transform martensitically, cold sprayed NiTi coatings suffer from poor adhesion to the substrates. In this work, we report on first ever successful fabrication of thick NiTi coatings (100-300 μm) that display functional thermomechanical properties and simultaneously show very good adherence to the substrate. We used high velocity air fuel thermal spray method to fabricate NiTi coatings deposited on mild steel using four different sets of processing parameters. Chemical composition, porosity, microstructure, phase transformation and functional thermomechanical properties of the NiTi coatings were evaluated. Although the coatings contain inhomogeneous microstructure, voids, oxide particles, high density of dislocation defects and internal stress, they undergo martensitic transformation upon cooling and/or mechanical loading. As sprayed NiTi coatings need to be annealed to display functional thermomechanical properties. Despite their limited tensile strength, the coatings displayed thermal actuation in 3-point bending tests and shape memory effects in nanoindentation and scratch tests.

Present address of Mohit Chandra: Neutron Scattering Division, Oak Ridge National Laboratory, Oak Ridge, Tennessee 37831, USA

Corresponding Author: Sneha Samal, Email:samal@fzu.cz

## 1. Introduction

NiTi shape memory alloys have already been widely used in engineering applications in medical devices, automotive aerospace, robotics, civil engineering and smart structures exploiting their unique superelastic, shape memory, and actuation properties derived from martensitic transformation (MT) supplemented by excellent corrosion resistance and biocompatibility [1-2]. Bulk NiTi alloys are conventionally produced by arc or induction melting with follow-up hot working and subsequently machined to the final dimensions. The melting/casting/forming approaches have their own advantages and drawbacks such as problems with assuring acceptable content of impurities, desired chemical composition and homogeneity and texture. Bulk NiTi components are therefore expensive due to the high material and processing costs.

Recently, with fast growth of powder metallurgy, NiTi alloys became available in a form of atomized powder. This opened space for wide range of new processing technologies such as laser powder bed fusion additive manufacturing or spark plasma sintering [3-7]. Besides that, it enabled deposition of thick NiTi coatings on the surface of bulk engineering components by thermal spraying. Thermal spray NiTi coatings deposited on the surface of engineering components modify their friction, wear, impact and corrosion resistance [8-10].

Various advanced plasma spray methods, such as Atmospheric Plasma Spray (APS), Vacuum Plasma Spray, Radio Frequency Inductively Coupled Plasma (RF-ICP) were used to fabricate such NiTi coatings [11]. However, these trials were largely unsuccessful due to the melting of NiTi particles which makes it difficult to fabricate chemically homogeneous single phase NiTi coatings. More promising are advanced thermal spray processes, such as High Velocity Oxygen Fuel (HVOF), high velocity air fuel (HVAF) and Cold Spray methods [12-14], used to apply dense, protective metallic or carbide layers onto surfaces, parts, and structures, which do not melt NiTi particles. There are differences among various thermal spray deposition techniques (see table S in Supplementary Materials). HVAF is a "warm spray" process that is cooler than HVOF, but hotter than Cold Spray.

Common problems of advanced thermal spray processes originate from excessive amount of oxygen present in HVOF, HVAF, and APS fabricated coatings. It gives rise to formation of titanium oxide phases [15-17] and affects porosity commonly found also in the HVAF fabricated NiTi coatings. Most of the articles reporting on deposition of thick NiTi coating by thermal spray methods deal with adhesion, microstructure characterization and evaluation of hardness, as common for hard coatings [18-23]. Very few articles report on evaluation of thermomechanical functional behaviour of NiTi coatings.

We have been exploring possibility of using HVAF thermal spray process to deposit thick NiTi coatings. HVAF yields a deposition efficiency of 60-85 % for metal powders. The HVAF process runs on a fuel gas using mixture of propylene or propane with compressed air. Nominally, metallic coatings fabricated by HVAF shall be less prone to internal oxidation than coatings prepared by HVOF and more resistant to porosity than cold sprayed coatings. Although we are concerned by both, we pay special attention to internal oxidation, which is of extreme importance for NiTi, the functional properties of which are strongly composition dependent.

In this work, we report results of pioneering research focusing on fabrication of thick NiTi metallic coating deposited on steel substrate by HVAF method using different processing parameters and characterization of their chemical composition, porosity, microstructure, phase transformation and functional thermomechanical properties.

**2**. **Experimental Methods**

The fabrication of NiTi coatings was performed by HVAF spraying using M2 nozzle 3L4C (210 mm long, 24 mm exit diameter, 17 mm throat diameter) and M3 nozzle 4L4C (254 mm long, 25.5 mm exit diameter, 18.9 mm throat diameter) from UniqueCoat, Richmond, USA. Further details regarding the M2 and M3 HVAF guns are available from the reference [24]. The mild steel (25.4 mm in diameter and 6 mm in thickness) was used as the substrate. The steel substrates were grit-blasted using alumina particles to achieve a surface roughness (Ra) of 3 μm, promoting mechanical interlocking and adhesion of the sprayed coating.

The phases of coating were detected at room temperature by using an X'Pert PRO θ-θ powder diffractometer (Malvern, UK) using Bragg–Brentano geometry at 40 mA and 35 kV with Kα Co radiation (average wavelength $\lambda = 0.1790$ nm), a focus-slit distance of 100 mm, and a goniometer radius of 240 mm. The data were measured in the 2θ range of 10–150° for powder and coatings, using a step size of 0.013°, a scan step time of 1.4 s, and a fixed divergence slit size of 0.5°. The microstructure of the powder and coatings was investigated by Scanning Electron Microscope (SEM, Tescan FERA 3). The cross-sectional surface of coating layers on the substrate of mild steel was prepared by cutting the coating with the substrate using an electric discharge machine. The samples were mounted on Bakelite with a follow-up metallographic polishing using an automatic grinder and polisher with up to 1 μm diamond paste.

Energy-dispersive X-ray spectroscopy (EDS) was carried out using the EDAX system (EDAX, Ametek Inc., Mahwah, NJ, USA) with an Octane Super 60 $mm^2$ detector to determine the chemical composition from substrate towards coating. The phase and grain orientation distribution analyses were performed using the electron back-scattered diffraction method (EBSD) using the EDAX DigiView V camera and system (EDAX, Ametek Inc., Mahwah, NJ, USA). A voltage of 20 kV, a current of 1–2 nA, and a working distance of 15 mm were chosen for analysis. The EBSD samples were metallographically prepared by the modified procedure for nickel (ASTM C-56 recipe). The colloidal silica was used in the last polishing step, and the surface was finished by etching in Kroll's reagent. The porosity within the coating was estimated via Image J software by using SEM cross-section images of the coating samples.

Martensite variant microstructures and oxide particles were analyzed by transmission electron microscopy (TEM) using microscopes Talos F200X G2 and FEI Tecnai TF20 X-Twin equipped with a field emission gun operating at 200 keV, a double tilt specimen holder, and an EDS detector. TEM foils were prepared by twin-jet electro-polishing using a Struers TenuPol-5 system inside a solution of 5 vol% perchloric acid and 95 vol% ethanol at −30 °C, followed by ion-beam thinning using a Gatan 691 system. Bright-field (BF), Dark-field (DF), high-angle annular dark-field (HAADF), and high-resolution TEM (HRTEM) imaging modes were employed in TEM analyses.

Phase transformation temperatures were detected for coating and annealed coatings (500 °C, 1 h, air medium) by Differential Scanning Calorimetry (DSC 25, TA Instruments, New Castle, DE, USA) at the rate of 5 ºC/min from – 100 °C to +100 °C in a nitrogen environment. Small samples were measured in an Al crucible (weights from 13.3 mg to 22.8 mg), taken for DSC measurement.

Mechanical testing was performed on coating samples in 3-point bending mode at a constant load of 500 mN at thermal cycles from – 60 º C to + 100 ºC by thermo mechanical analyzer (TMA, using TMA PT 1600 (Linseis, Selb, Germany) with a standard ASTM E831 [16]. The tensile tests were performed by using a custom-made tensile tester MITTER (Miniature Testing Rig), equipped with an environmental chamber, electrically conductive grips, a load cell, a linear stepping motor, and a position sensor. The testing device enabled us to control temperature and stress and measure the strain of the samples. The tests were conducted at room temperature, with a constant strain rate of 0.05 %/s to reach the maximum load until the coating fracture. The functional property of the coating was tested by the Discovery DMA 850 tester (TA Instruments, New Castle, USA) at thermal cycles ranging from −60 to +100 °C with a rate of 5 °C/min. In this test, a sample of dimension 25 × 5 × 0.3 $mm^3$ was heated to 100 °C first, then cooled to room temperature, loaded to 300 MPa in bending mode in a thermal cycle from cooling and heating. The multiple cycles were performed on the bending mode to examine the functional shape memory effect.

The multicycle nano indentation of the sprayed coating samples was carried out by using conospherical tip of diameter 10 micron in 4 x 4 array on the surface of the samples at various loads from 50 -200 mN using FEMTO tool in-situ SEM nano indenter. The aim is to analyze the shape memory effect and the corresponding elastic and plastic recovery. Also, we observe the cyclic stability of recovered and unrecovered depth until $40^{th}$ cycle of multi-cycle nanoindentation. Coating specimens for scratch measurements with size of 10 x 10 x 0.25 mm were cut from the substrate. To remove pre-existing surface oxide layers, sandpaper was used for rough grinding of the samples, then polished with diamond paste and electrochemical polishing in 5 % perchloric acid at 40 V was employed to prepare a stress-free surface. Progressive-load scratch tests were performed on the samples at fixed load using the using a fully calibrated NanoTest P3 instrument (MicroMaterials, Wrexham, Wales) with a spherical indenter of 10 µm radius, in the "scratch mode" configuration, with an indenter located perpendicularly to the sample, while the sample moved along the vertical axis. The progressive scratch test consisted of three passes. The first pass scans the initial topography at a maximum load of 0.05 mN, followed by the loaded scratch pass and the final topography pass performed at the same load as the initial topography. The scratch pass consists of surface scanning with a constant topography load of 0.05 mN over the first 50 µm of the track; subsequently, the load is linearly ramped over the next 400 µm of the track up to the selected maximum load (500 mN).

## 3. Results and Discussion

### 3.1. Pre-alloy NiTi Powders

Pre-alloy gas atomized powder of nominal composition $Ni_{50}Ti_{50}$ (at. %) with 99.5 % purity obtained from American Element (AE, MERELEX CORPORATION, LOS ANGELES, CA 90024, USA) was used to fabricate the HVAF NiTi coatings. The powder consists of spherical particles with size ranging from 5 to 55 µm (Fig. 1b). The powder displays martensitic transformation on cooling ($M_s$ = 39 ºC) and heating ($A_f$ = 62 ºC) as determined by DSC thermal cycling (Fig. 1c). At room temperature, the NiTi powder thus can be either austenitic or martensitic. The small DSC peaks are assumed to be due to chemical heterogeneity, possibly introduced during the atomization. Analysis of chemical composition by EDX in SEM gave higher nickel content (Fig. 1d,e) compared to the nominal composition. Nevertheless, given the 2% uncertainty of the method, the result is acceptable, though $A_f$ temperature 62 ºC corresponds more to the nominal composition.

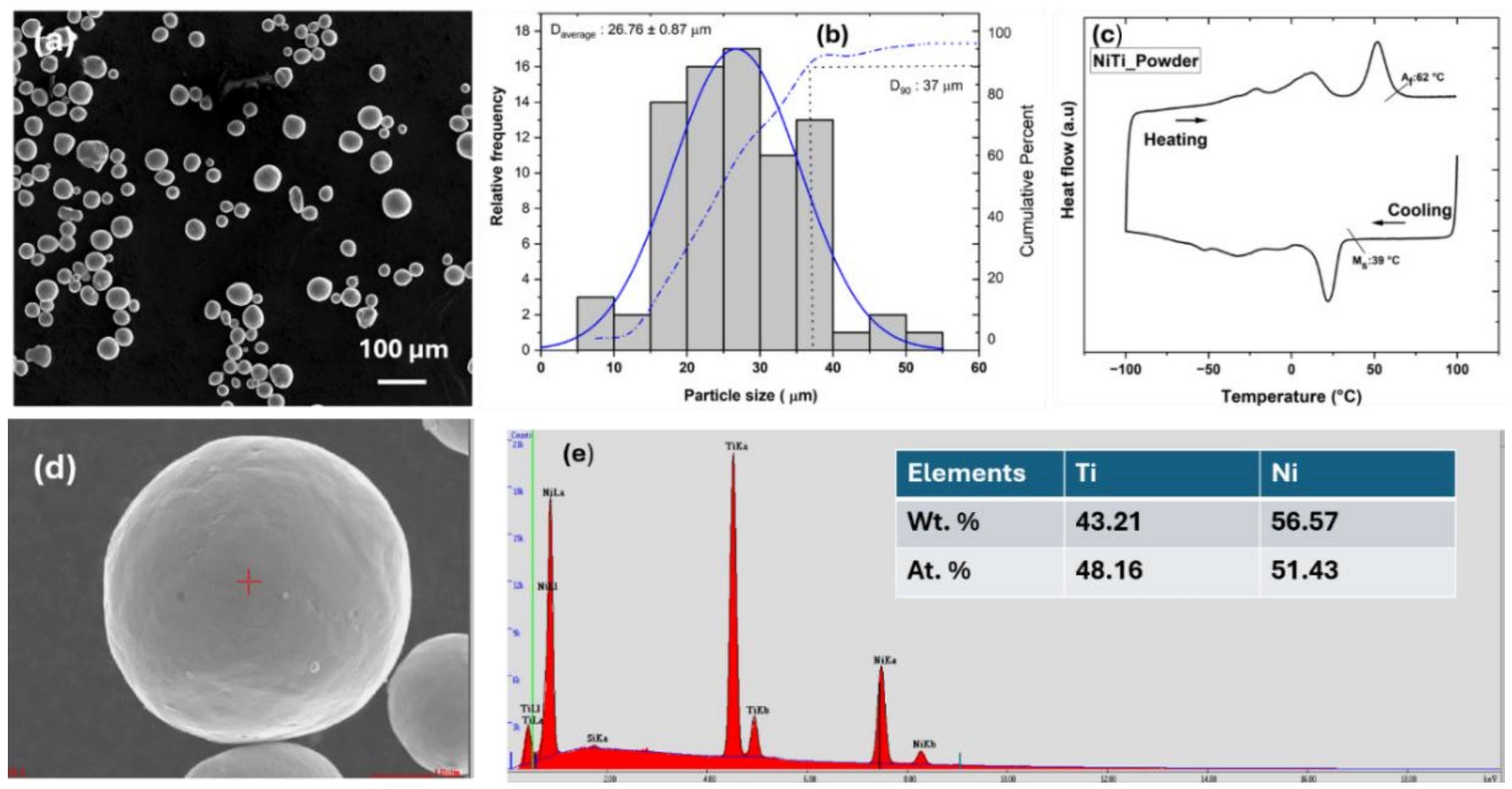

**Fig. 1:** NiTi powder characteristics a) Size and shape of the NiTi particles characterized by SEM, b). Histogram of powder size distribution ($d_{avg}$: 26.76 µm, $d_{90}$: 37 µm), c) DSC scan of the NiTi powder performed to evaluate transformation temperatures ($M_s$ =39 ºC, $A_f$=62 ºC) d) red mark in the powder shows the location for chemical analysis, e) Detail of particle used to analyze chemical composition by EDX, f) Results of the analysis of chemical composition.

### 3.2. HVAF fabrication of NiTi

Four different NiTi coatings were fabricated using different sets of processing parameters (Table 1). Main goal of the fabrication trials was to find out whether and how the characteristics and mechanical properties of as-sprayed NiTi coatings are sensitive to variation of processing parameters. The differences consist in different nozzles, air pressure and thickness of the HVAF coatings. Another goal of this work was to identify proper experimental methods to be used for characterisation of fabricated NiTi coatings and evaluation of their functional mechanical properties. The methods are very important since the the HVAF process needs to be optimized with respect to the results of these methods.

It was found that the as-sprayed coatings need to be subjected to post processing heat treatments to display functional mechanical properties. This opens new possibilities, since the quality of fabricated coatings needs to be optimized also with respect to the temeperatre and tine of the post processing annealing treatment. Therefore, the as-sprayed NiTi coatings were annealed at 500 °C for 1 hour to bring them into the state, in which they undergo MT and display functional mechanical properties. Nevertheless, we characterized systematically the as-sprayed NiTi coatings, because the as-sprayed state is the original well defined state.

**Table 1** represents the HVAF spray deposition parameters for the preparation of coating samples 1-4.

| | **Sample 1** | **Sample 2** | **Sample 3** | **Sample 4** |
|---|---|---|---|---|
| **Gun Type** | M2 | M2 | M3 | M3 |
| **Nozzle** | 3L4c (nozzle 1) | 3L4c (nozzle 1) | 4L4c (nozzle 2) | 4L4c (nozzle 2) |
| **Chamber** | M3/2 | M3/2 | M3/1 | M3/1 |
| **Injector** | ti64x | ti64x | ti64x | ti64x |
| **Fuel type** | Propane | Propane | Propane | Propane |
| **Fuel 1 ( PSI)** | 100 | 100 | 100 | 100 |
| **Fuel 2 ( PSI)** | - | - | 105 | 105 |
| **Air ( PSI)** | 100 | 100 | 108 | 108 |
| **SOD (mm)** | 200 | 200 | 300 | 300 |
| **Feed rate (g/min)** | 50 | 50 | 50 | 50 |
| **Carrier gas (LPM)** | 40 | 40 | 40 | 40 |
| **Coating thickness (μm)** | 100 | 300 | 100 | 260 |

(carrier gas – 40 lpm; powder feed rate – 50 g/min were kept constant for all runs), SOD : stand-off distance, Carrier gas ( LPM) : Liters per minute

### 3.3. Characterisation of NiTi coatings

#### 3.3.1. Phase compositions

Besides evaluating the chemical composition, we need to know whether the NiTi coating transforms martensitically and what is the phase composition at desired temperature. For HVAF NiTi coatings to be used at environmental temperatures, desired temperature is room temperature. Fig. 2 compares X-ray diffraction pattern of NiTi powder recorded at room temperature with diffraction patterns of NiTi coatings 1-4. While the NiTi powder contains mixture of austenite and martensite, all four NiTi coatings are austenitic. This can be rationalized by considering suppression of $M_s$ by mechanical deformation during the spraying and/or change of chemical composition due to internal oxidation (NiTi matrix becomes enriched in Ni due to depletion of Ti forming Titanium oxides (Chapter 3.3.3). The phase fractions and lattice parameters of phases determined by quantitative analysis by Rietveld refinement of the experimental XRD patterns are listed in Table 2 [25].

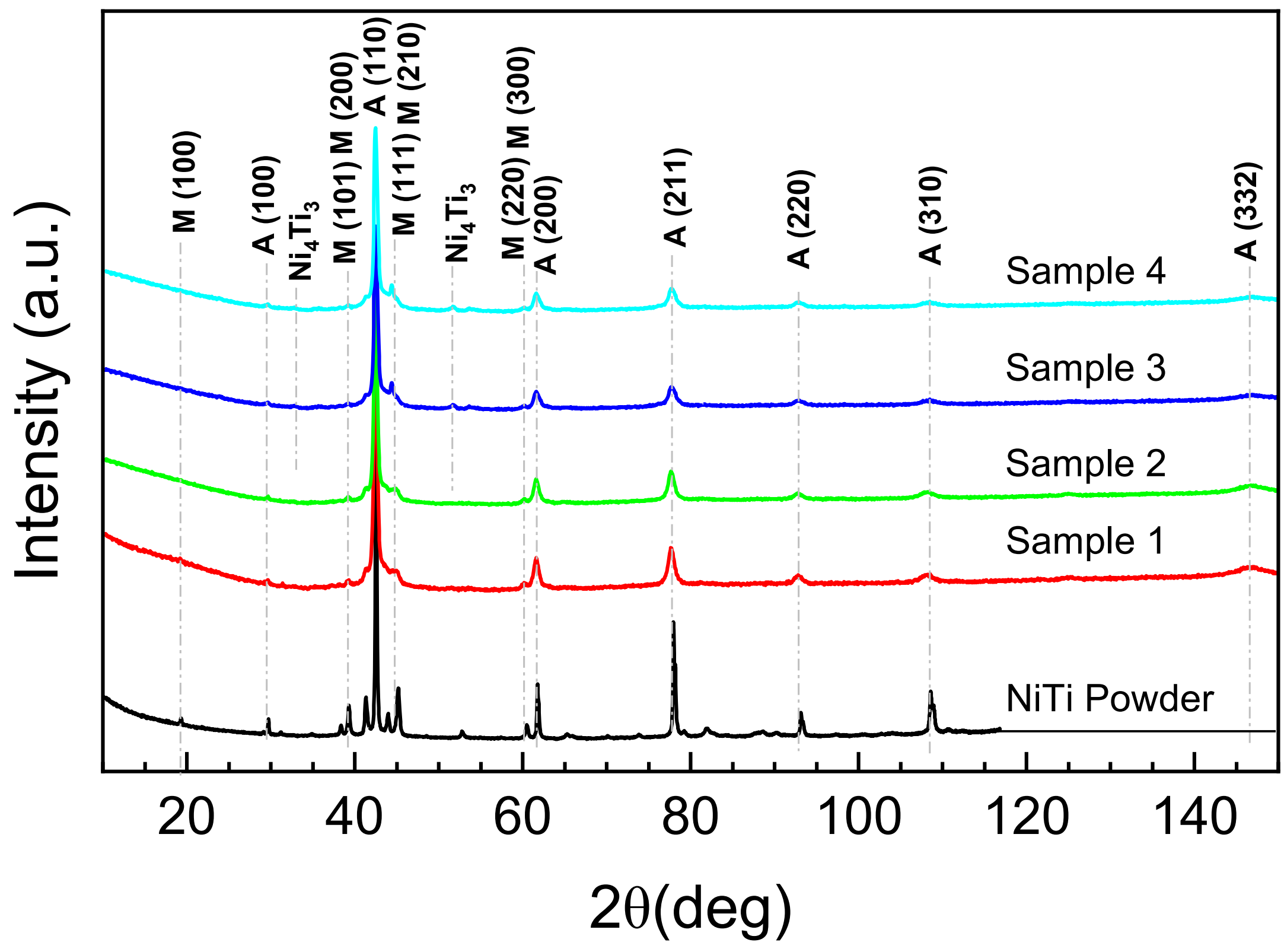

**Fig. 2:** X-ray diffraction patterns of the NiTi powder and HVAF NiTi coating 1-4 at room temperature.

**Table 2**. Results of the quantitative phase analysis of NiTi powder and HVAF NiTi coatings 1-4.

| Sample | Phases | Lattice Parameters | Quantitative (Wt. %) |
|---|---|---|---|
| NiTi Powder | Austenite (Cubic, Pm-3m) | a=3.010(5) | 64 |
| | Martensite (Monoclinic; P21/m) | a=4.619(8)<br>b=4.149(8)<br>c=2.889(4)<br>β=96.67(9)° | 36 |
| Sample 1 | Austenite (Cubic, Pm-3m) | a = 3.010 (5) | 92.3 |
| | Martensite (Monoclinic; P21/m) | a = 4.619 (8)<br>b = 4.149 (8)<br>c = 2.889(4)<br>β = 96.67 (9)° | 7.7 |
| Sample 2 | Austenite (Cubic, Pm-3m) | a = 3.010 (5) | 89.43 |
| | Martensite (Monoclinic; P21/m) | a= 4.617(8)<br>b= 4.147(7)<br>c= 2.893(3)<br>β= 96.61(7) ° | 10.57 |
| Sample 3 | Austenite (Cubic, Pm-3m) | a = 3.009 (2) | 81.3 |
| | Martensite (Monoclinic; P21/m) | a= 4.630 (9)<br>b= 4.127 (5)<br>c = 2.917(4)<br>β = 96.07 (6)° | 15.1 |
| | $Ni_4Ti_3$ (Trigonal; R-3) | a=b=11.0 (1)<br>c=5.2(3)<br>γ = 120 ° | 3.5 |

| Sample 4 | Austenite (Cubic, Pm-3m) | a = 3.009 (2 | 80.4 |
|---|---|---|---|
| | Martensite (Monoclinic; P21/m) | a= 4.630 (9)<br>b= 4.127 (5)<br>c = 2.917(4)<br>β = 96.07 (6)° | 15.8 |
| | $Ni_4Ti_3$ (Trigonal; R-3) | a=b=11.0 (1)<br>c=5.2(3)<br>γ = 120 ° | 3.7 |

### 3.3.2. SEM analysis of microstructures

Although we did not evaluate the adhesion, we believe that the fabricated HVAF NiTi coatings form very strong interfaces with the substrate. We did not encounter any detachment in bending and/or tensile tests performed on coated substrate samples. Fig. 3 displays the cross-section images of HVAF NiTi coatings 1-4. The interfaces are irregular, partly because the substrate was roughened and partly because of the HVAF compaction process.

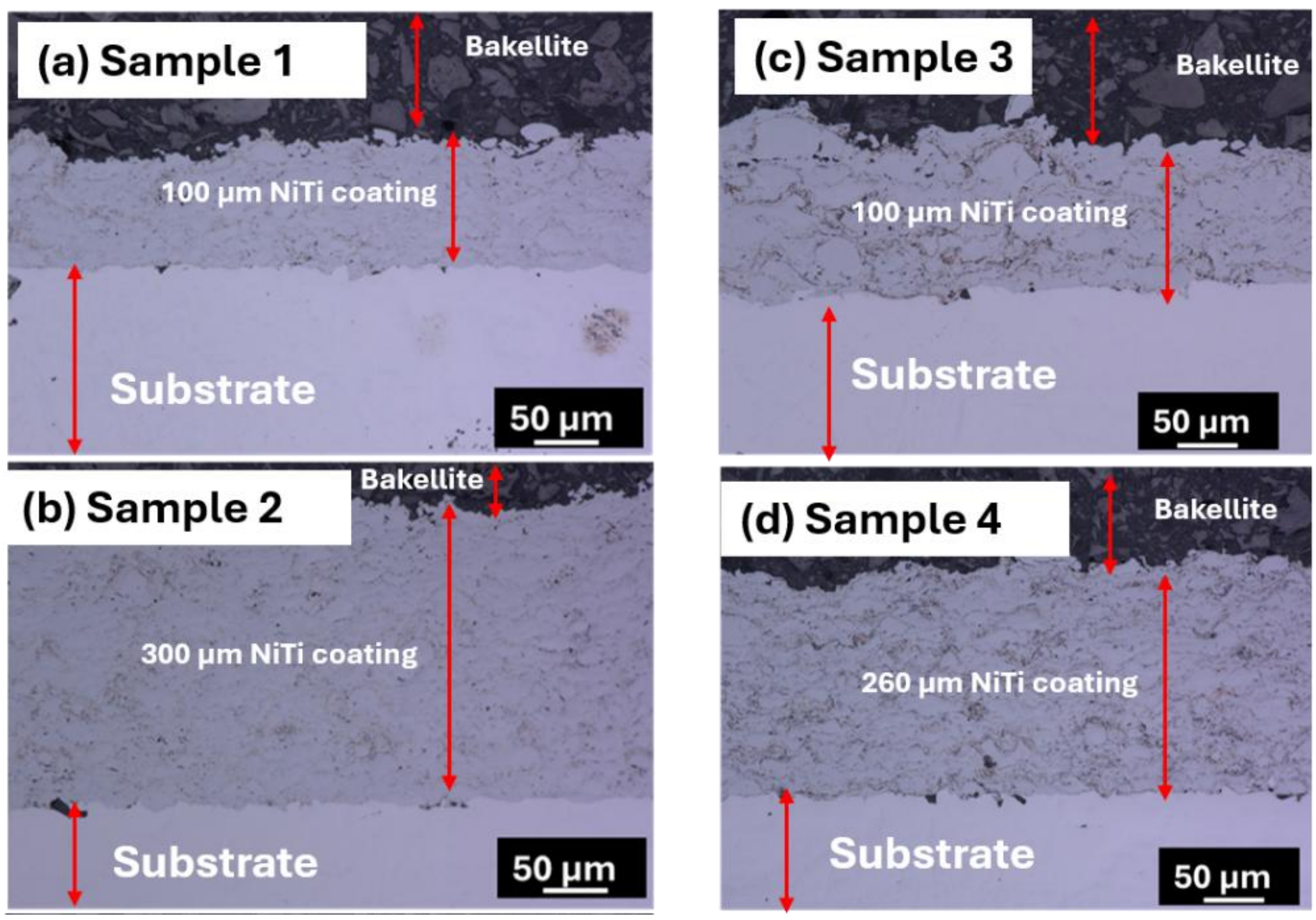

**Fig. 3:** Cross-section images through the substrates of HVAF NiTi coatings 1-4 of different thicknesses. The samples 1 and 2 were prepared from nozzle 1 and samples 3-4 were prepared from nozzle 2.

The fabricated NiTi coatings display significant porosity that depends on the used process parameters (Fig. 4, Table 3). Specifically, NiTi coatings prepared using nozzle 1 (samples 1,2) show higher porosity than NiTi coatings prepared from nozzle 2 (samples 3,4). Thicker coatings show higher porosity than thinner coatings. The mechanism responsible for creation of the pores and its link to the processing parameters is currently being investigated. Differences may be also due to increasing thickness.

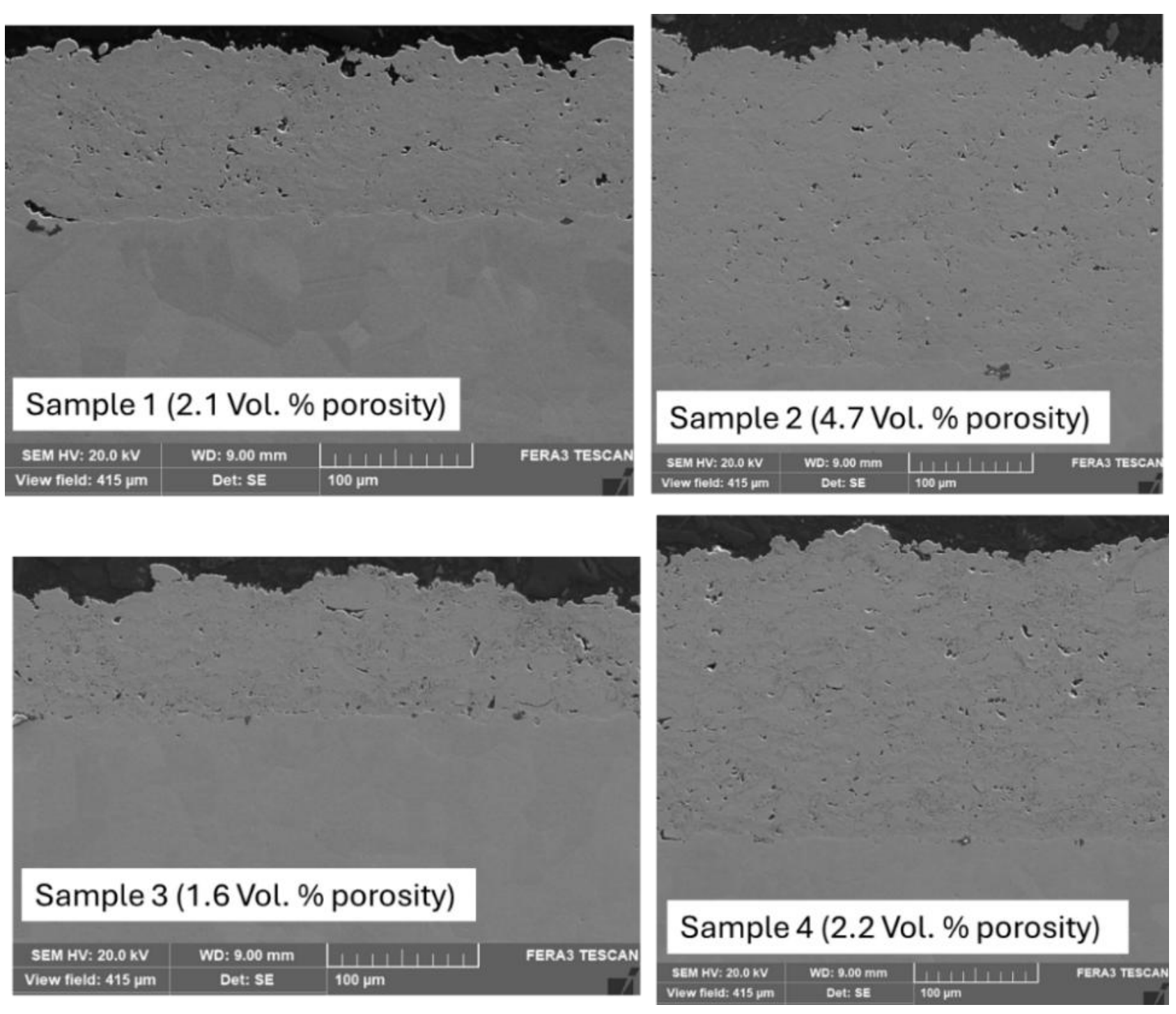

**Fig. 4:** Samples 1 and 2 are prepared from a nozzle with two different thicknesses of the coating (100 µm, 300 µm), showing porosity of 2.1 Vol. % and 4.7 vol. %. Samples 3-4 are prepared from a nozzle with two different thicknesses of the coating (100 µm, 260 µm), showing porosity of 1.6 Vol. % and 2.2 Vol. %.

**Table 3:** Porosity within the NiTi coatings 1-4 (vol. %)

| Coating | Total area | Porosity area | Vol. % of porosity |
|---|---|---|---|
| Sample 1 | 3924 | 84.9 | 2.1 |
| Sample 2 | 6312.2 | 296.0 | 4.7 |
| Sample 3 | 7393 | 119.3 | 1.6 |
| Sample 4 | 11104 | 252 | 2.2 |

Chemical composition of the HVAF NiTi coatings 1-4 was evaluated on the cross section by EDX (Fig. 5, Table 4). It is found that Ni content is lower in the NiTi coatings than in the powder. At the same time, chemical composition depends on the processing parameters. NiTi coatings 1 and 2 prepared using nozzle 1 show higher Ni content than NiTi coatings 3 and 4 prepared using nozzle 2. The results of the line scan analysis (Fig. 5a-d) prove macroscopic homogeneity of chemical composition across the coating thickness.

**Table 4.** Chemical composition of the NiTi powder and HVAF NiTi coatings 1-4 evaluated by EDX (average over the coating cross section)

| | Ni (wt. %) | Ti wt. (%) | Ni (at. %) | Ti (at.%) |
|---|---|---|---|---|
| Sample 1 | 55.3 | 44.2 | 50.2 | 49.1 |
| Sample 2 | 55.3 | 44.4 | 50.2 | 49.4 |
| Sample 3 | 54 | 45.6 | 48.9 | 50.6 |
| Sample 4 | 54.3 | 45.2 | 49.3 | 50.2 |

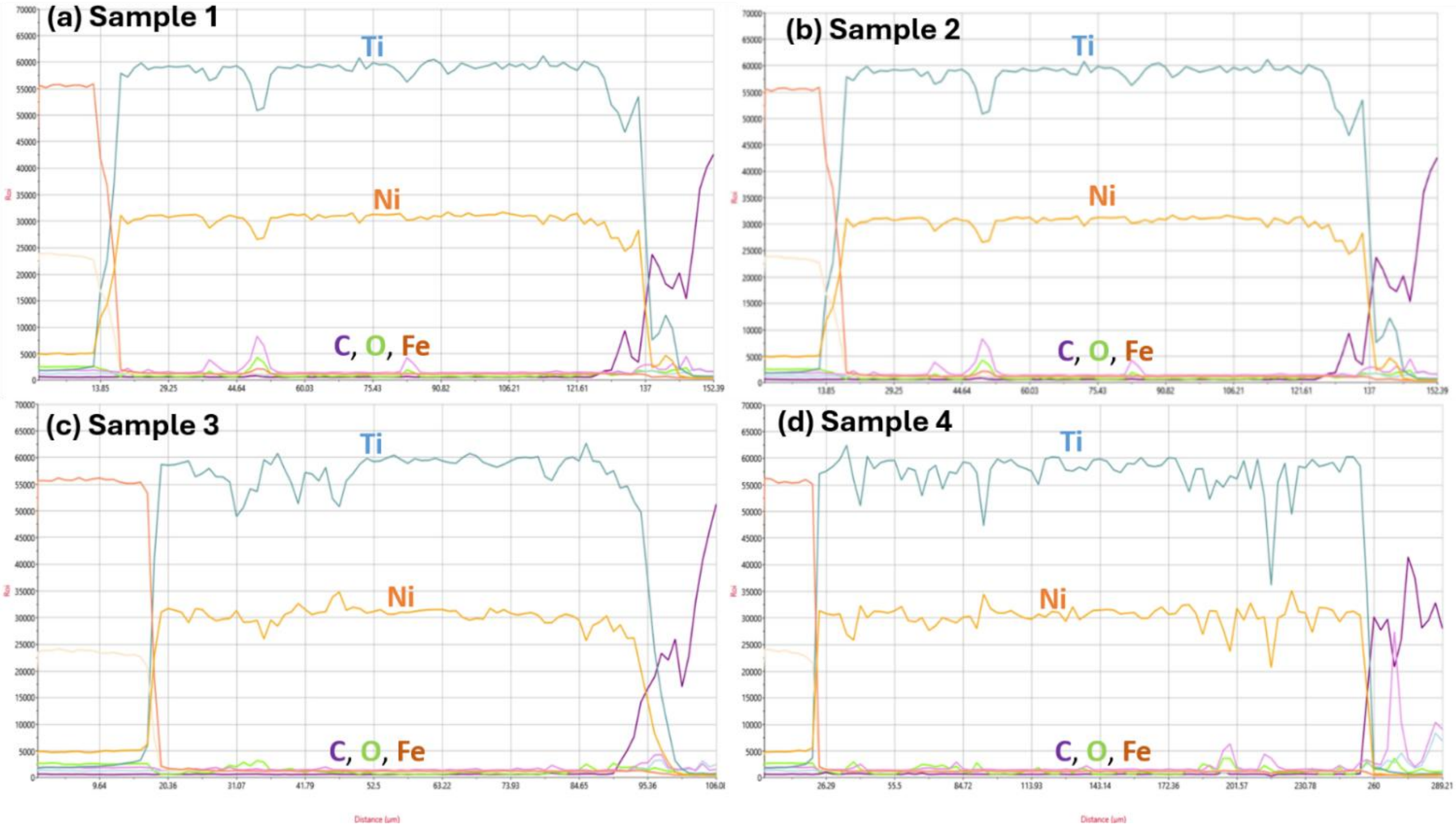

**Fig. 5:** Line scan analyses of the macroscopic homogeneity of chemical composition of the NiTi coatings 1-4 performed by EDX shows the presence of elements Ni, Ti, with negligible content of other elements such as C, O, Fe.

### 3.3.3. TEM analysis of microstructure

The microstructures of HVAF NiTi coatings 2 and 4 were analyzed by transmission electron microscopy (TEM) with the aim to evaluate the present phases, grains size and lattice defects (Fig. 6). Since the as-fabricated NiTi coatings (Fig. 6a,c) contain internal stress disturbing the electron diffraction contrast, we analyzed also NiTi coatings annealed at 500 °C for 1 hour (Fig. 6b,d), as common in NiTi technology. The internal stress in the as-sprayed coatings was relieved by the annealing only partially, density of dislocation defects decreased and grain boundaries became clearly visible.

While coating 2 is mainly austenitic, as evidenced by electron diffraction patterns (Figs. 6a4,b4) taken from SAED areas denoted by yellow circles, coating 4 contains mixture of austenite and martensite phases (Figs. 6d2,d4). and many $TiO_2$ oxides (Figs. 6c3,c4). Since the heated NiTi particles hitting the substrates are covered by thin (20-100nm) naturally grown $TiO_2$ oxide, there is excessive oxygen and Ti everywhere - homogeneously distributed within the coating. Results of the TEM analysis of microstructure (Figs. 6c3,c4) however shows very inhomogeneous distribution of Titanium oxides.

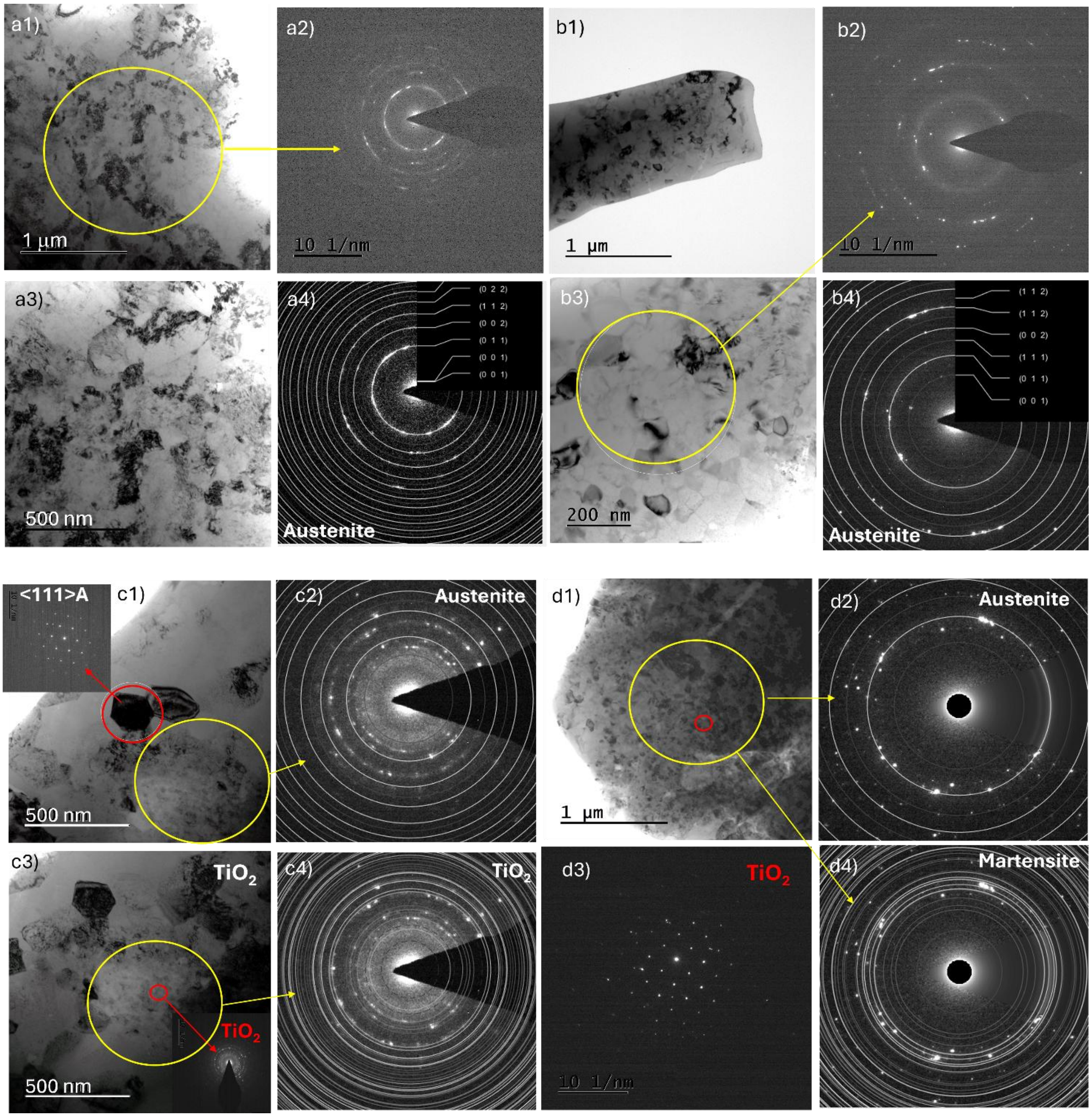

**Fig. 6**: Results of the TEM analysis of HVAF NiTi coating 2 (a,b) and coating 4 (c,d) in the as sprayed state (a,c) and after annealing at 500 ºC for 1 hour (b,d).

### 3.3.4. Martensitic transformation

Besides chemical composition, phase composition and microstructure, NiTi alloy is characterized by the occurrence and type of martensitic transformation and transformation temperatures. Whether MT takes place in NiTi powder as well as in as-fabricated and annealed NiTi coatings was evaluated by DSC thermal cycling (Fig. 7,8). DSC scans of the as-sprayed NiTi coatings (Fig. 7) do not show any clearly defined heat flow peaks when cooled to -125 °C. and heated up to 100 °C. This does not mean necessarily that they do not transform upon thermal cycling. It is more likely that the MT is spread over wide temperature interval due to the internal stress and/or local chemical heterogeneity which make the DSC peaks to become extremely broad. In addition, transformation to B19' martensite can be partially suppressed to temperatures below -125 °C. This can be understood based on the assumption that NiTi particles become plastically deformed and oxidized while the NiTi coatings form during the thermal spraying process. Although the DSC curves in Fig. 7 resemble DSC curves of cold-worked NiTi [26-28], the austenitic microstructure of as-sprayed NiTi coatings is different from the microstructure of cold drawn NiTi wires.

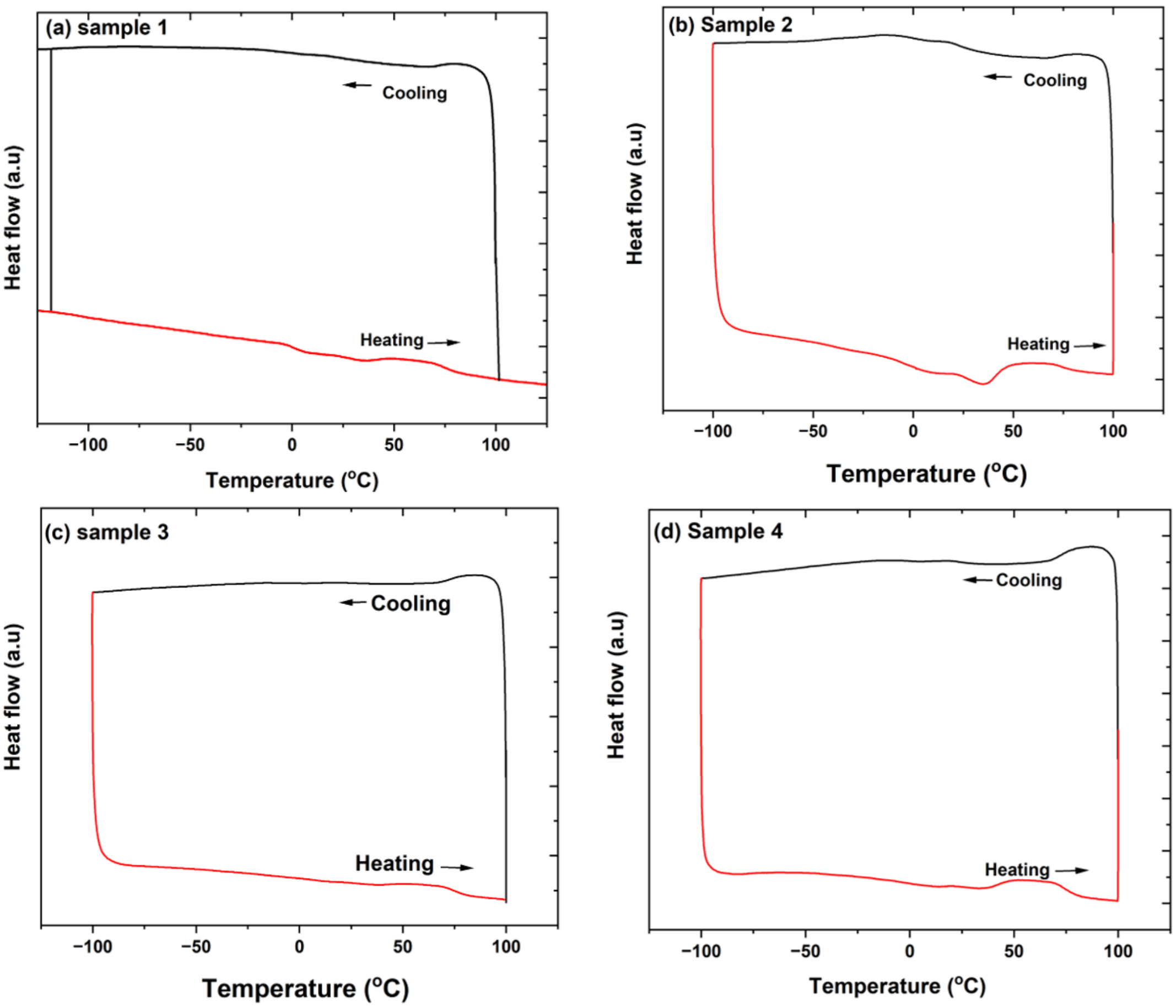

**Fig. 7:** Results of DSC thermal cycling of as-fabricated HVAF NiTi coatings 1-4 (a-d) performed to evaluate transformation temperatures. The thermal cycling was performed in the temperature range of −100 ºC to +100 ºC with a cooling/heating rate of 5 ºC/min.

However, when the as-sprayed NiTi coatings were annealed at 500 °C for 1 h, results of DSC thermal cycling changed (Fig. 8). DSC peaks characterizing MT proceeding on cooling and heating appeared on DSC scans. Compared to the NiTi powder which undergoes B2-B19' MT on cooling and heating (Fig. 1c), annealed NiTi coatings (Fig. 8) display B2-R-B19' MT (transformation temperatures are listed in Table 5). All four NiTi coatings show B2-R transformation around room temperature. The DSC peaks corresponding to transformation to the B19' martensite are, however, rather small and broad (compar6 to the DSC peaks of NiTi powder in Fig. 1f). This suggests that the internal stress was not completely relieved by the applied annealing and that there can be local chemical heterogeneity in the microstructure that causes the MT to proceed in wide temperature range [29]. Further investigations of the peculiar microstructure in as-sprayed as well as in annealed NiTi coatings is necessary to understand the martensitic transformations proceeding in it.

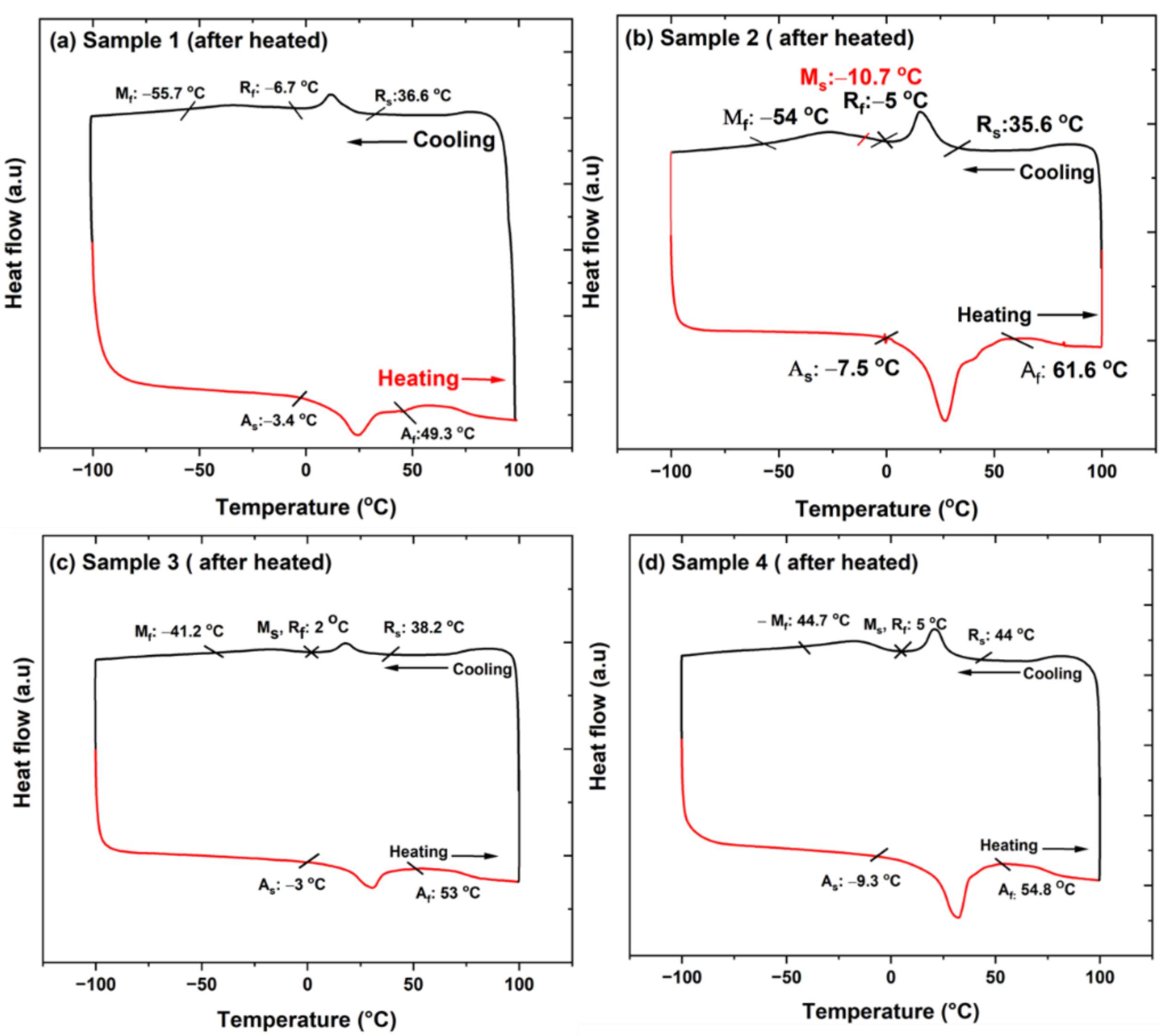

**Fig. 8:** Results of DSC thermal cycling of HVAF NiTi coatings 1-4 annealed at 500 ºC for 1 h (a-d) performed to evaluate transformation temperatures. The thermal cycling was performed in the temperature range of –100 ºC to +100 ºC with a cooling/heating rate of 5 ºC/min.

**Table 5.** Transformation temperature of the phases of HVAF NiTi coatings 1-4 evaluated by DSC tests.

| Sample | Austenite start ($A_s$, °C) | Austenite finish ($A_f$, °C) | R-phase start ($R_s$, °C) | R-phase finish ($R_f$, °C) | Martensite start ($M_s$, °C) | Martensite finish ($M_f$, °C) |
|---|---|---|---|---|---|---|
| Sample 1 | –3.4 | 49.3 | 36.6 | -6.7 | - | -55.7 |
| Sample 2 | – 7.5 | 61.61 | 35.6 | –5 | –10.7 | – 54 |
| Sample 3 | –3 | 53 | 38.2 | 2 | 2 | – 41.2 |
| Sample 4 | –9.3 | 54.8 | 44 | 5 | 5 | – 44.7 |

### 3.4. Functional thermo-mechanical properties

Finally, HVAF NiTi coatings shall display unique functional mechanical properties derived from the MT that need to be evaluated experimentally. Since the fabricated coatings are rather thick (100-300 µm), they could be separated from substrates by spark cutting. The detached coatings were polished and subjected to dilatometry tests (Fig. 9), mechanical tests in tension (Fig. 10) and thermomechanical loading tests in bending (Fig. 11).

Since separation of the NiTi coating from the substrate is quite laborious and testing of detached coatings is not really relevant for their intended applications, we need to develop and use experimental methods capable of evaluating the mechanical properties of NiTi coatings without detaching them from substrates. These methods will be used later when optimizing the HVAF process to fabricate NiTi coatings having desired functional thermomechanical properties. In this work, we performed scratch tests (section 3.4.2) and nanoindentation tests (section 3.4.3) to evaluate mechanical properties of the NiTi coatings.

#### 3.4.1. Dilatometry

The dilatometry experiments were performed on as-fabricated HVAF NiTi coatings 1-4 subjected to 3-point bending thermal cycling test (five cooling-heating cycles from – 60 º C to + 100 ºC under 500 mN force). Based on the results of dilatometry tests, it can be determined whether the material displays conventional thermal dilatation or hysteretic responses due to martensitic transformation. The recorded nonlinear hysteretic responses (Fig. 9) indicate that MT proceeds in thermally cycled coatings in wide temperature range, even if the DSC test did not show heat flow peaks. Nevertheless, the samples responded by very small bending strains suggesting that massive MT was suppressed below -60 ºC. The results of dilatometry experiments thus revealed show that the as-sprayed NiTi coating transform martensitically and that the MT depends on HVAF processing parameters.

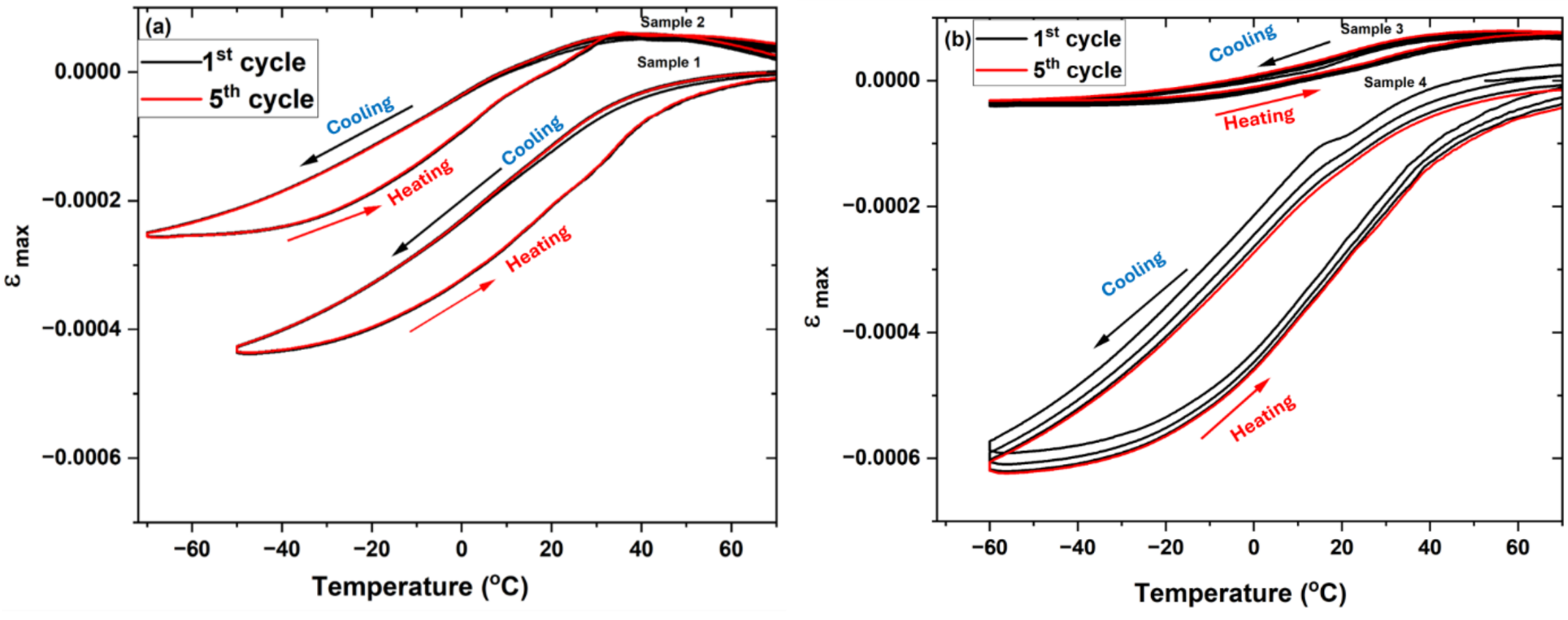

**Fig. 9:** Results of thermomechanical cycling (dilatometry test) of a) HVAF NiTi coatings 1 and 2 ( nozzle 1) (a) and b) HVAF NiTi coatings 3 and 4 (nozzle 2). The graphs show the dependence of maximum strain ($\varepsilon_{max}$) on temperature in 3-point bending test at a relatively low constant force 500 mN (~30 MPa stress on the surface).

#### 3.4.2. Young's modulus and tensile strength

Next, tensile tests at room temperature were performed on as-sprayed HVAF NiTi coating separated from substrates to evaluate Young's moduli and tensile strengths (Fig. 10, Table 6). It is found that Young's moduli are lower (~45 GPa) than expected for B2 austenite (~70 GPa) and tensile strength is very low (~200MPa). The coatings fractured in tensile test in elastic range at tensile strains ~0.4%. Although these results were rather disappointing, we need to consider that the as-sprayed NiTi coatings contain pores, oxides and internal stresses. Optimization of the HVAF processing will be necessary to increase the tensile strength.

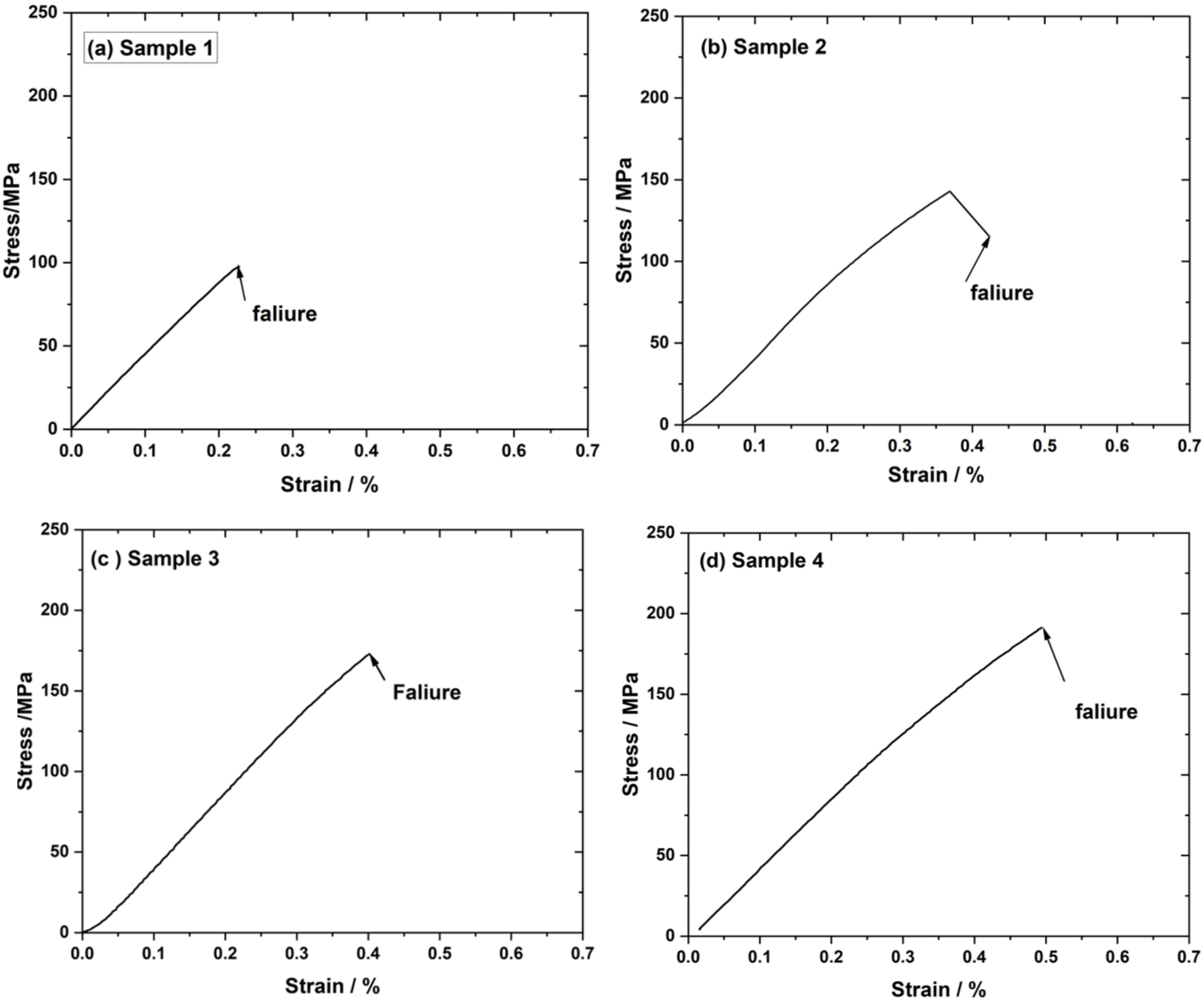

**Fig. 10.** Results of tensile tests at room temperature performed on as-sprayed HVAF NiTi coating separated from the substrates (a-d).

To find out whether the NiTi coatings display functional properties despite their low tensile strength, we performed 3-point bending thermomechanical loading tests in DMA tester. However, the as-sprayed NiTi coatings showed only unrecovered plastic strain developing upon first cooling in the bending test, but no strain reversal upon the subsequent heating and further cycling. Therefore, the experiment was repeated on NiTi coating 4 annealed at 500 °C for 1 hour (Fig. 11). The results confirm the presence of MT upon thermal cycling. The bending strain starts to increase at room temperature at which the coating 4 undergoes B2-R transformation (Fig. 8d). The shape of the strain-temperature hysteretic curve, however, is not standard due to the peculiar microstructure (Fig. 6d) in the annealed NiTi coating.

Table 6 : Young's moduli and tensile strengths of as fabricated NiTi coatings 1-4 determined from tensile tests at room temperature

| Coating | Youngs Modulus ( GPa) | Strength of coating (MPa) |
|---|---|---|
| Sample 1 | 43.9 | 96 |
| Sample 2 | 46.8 | 141.7 |
| Sample 3 | 45.3 | 170.9 |
| Sample 4 | 45.8 | 193.6 |

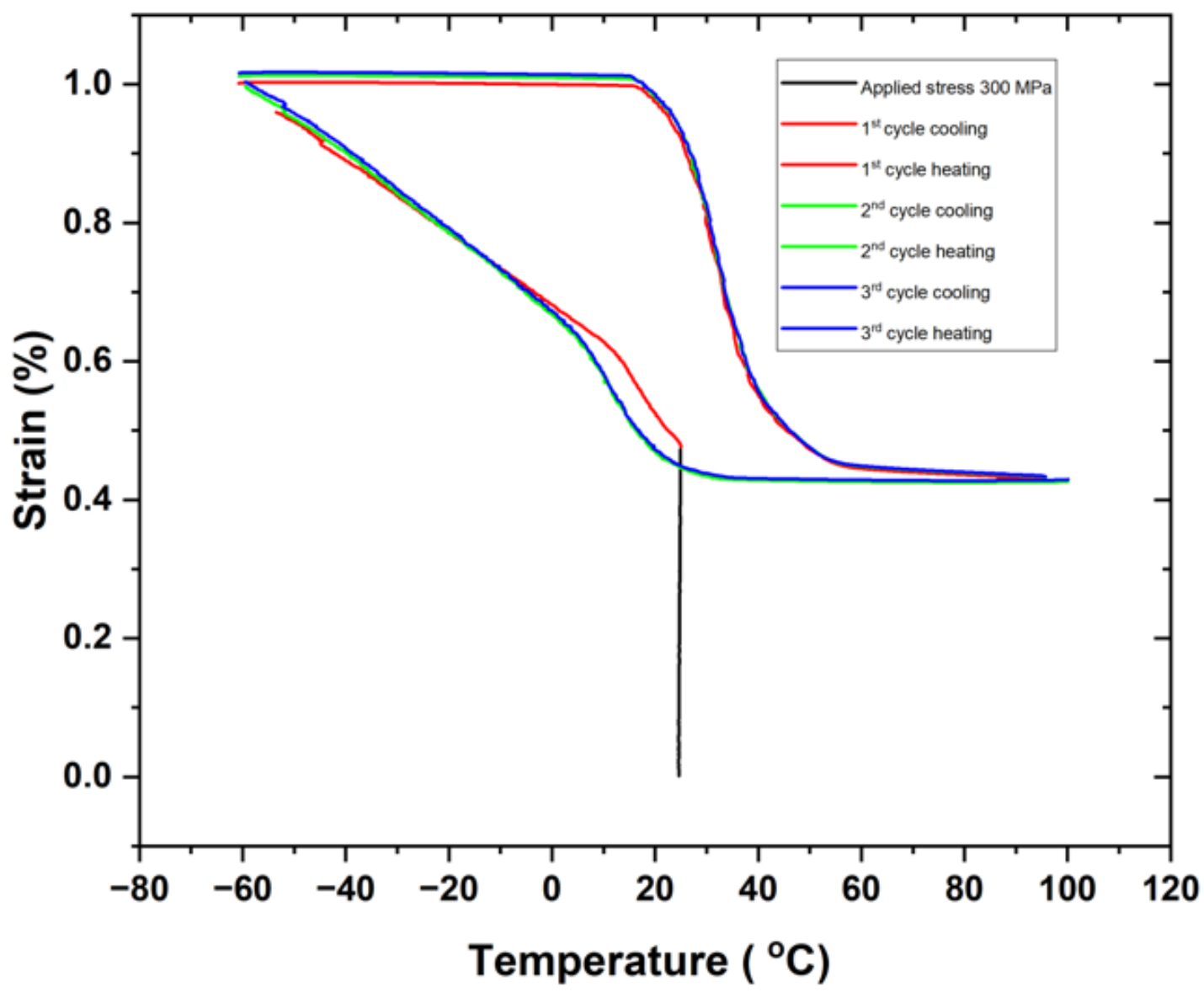

**Fig. 11.** Results of 3-point bending test (thermally cycle under 300 MPa stress) on NiTi coating 4 annealed at 500 °C for 1 hour.

### 3.4.3 Mechanical properties evaluated by scratch tests

Although the experimental results obtained in experiments on NiTi coating separated from the substrate are very informative, mechanical properties of the coatings need to be evaluated by experimental methods, which do not require to separate them from the substrates. We present and discuss in this article only selected results of scratch tests and nanoindentation tests performed on as-fabricated NiTi coatings 1-4.

Fig. 12 displays results of the scratch tests (progressive loading up to 500 mN) on the steel substrate and as-fabricated NiTi coatings samples 2 (300 µm, nozzle 1) and sample 4 (260 µm, nozzle 2). Initial topography (yellow), scratch depths (blue), and final topography (red) response for three scratch lines are shown. In general, a jagged pattern of the curve indicates roughness on the surface. Therefore, the elastic recovery was estimated from linear fits (baselines) of the loaded scratch depth profile and the final topography profile, with the recovery value evaluated at the end point of the scratch track.

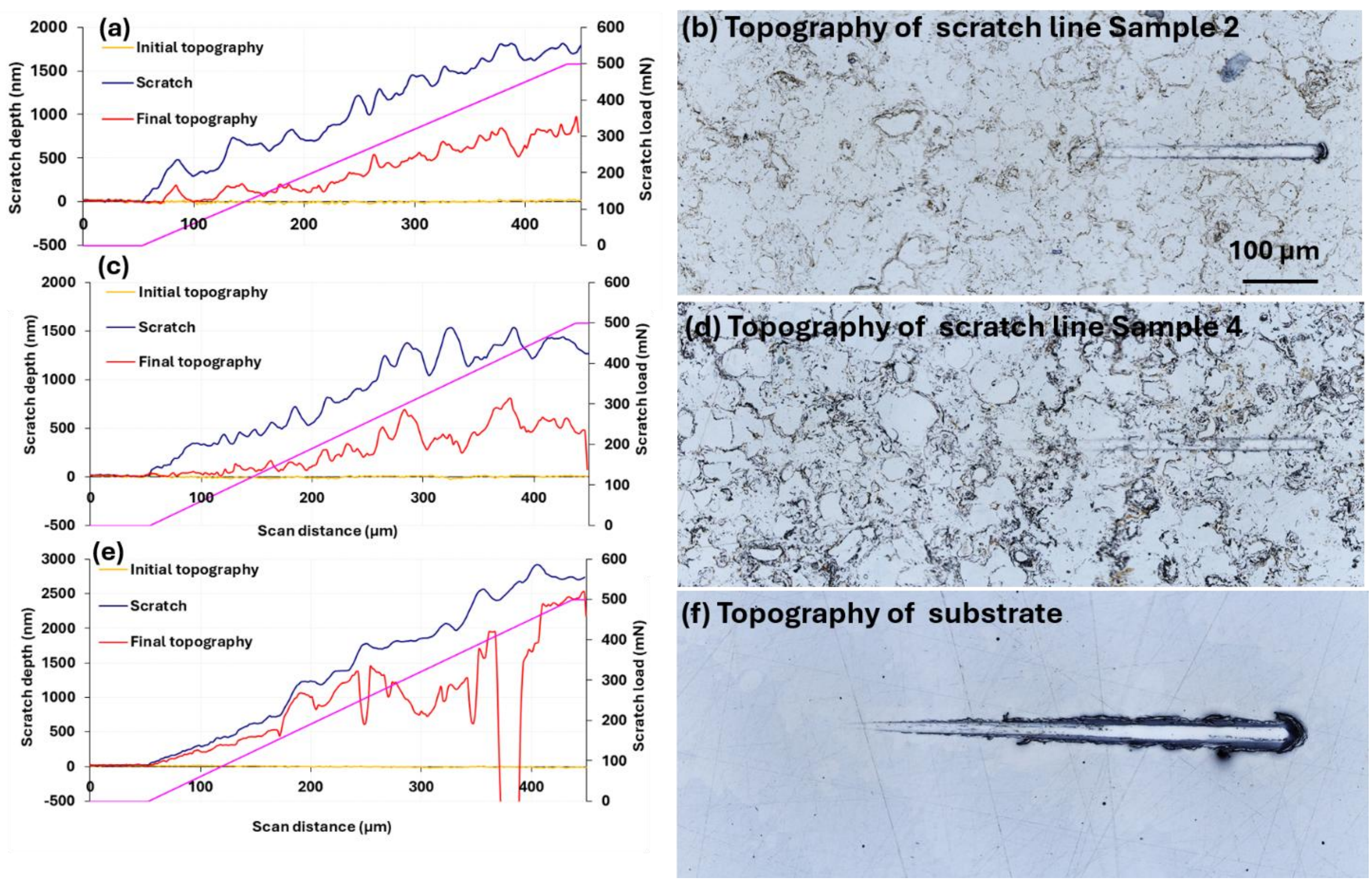

**Fig.12.** Progressive-load scratch tests performed on coating 2, 4 and mild steel samples up to maximum loads of 500 mN with a conospherical indenter of radius 10 µm. The depth changes are recorded and represented as the different phases of the test are marked in color, with yellow representing the initial topography, blue representing the load scratches, and red representing the final topographies (a-b). (a) Scratch test for HVAF coating ( a, c) and steel substrate (e). The image of the scratch line is shown in (b), (d) for coating and (f) steel substrate. The residual scratch line is observed on the coating surface and substrate after unloading from maximum load of 500 mN.

The substrate exhibits plastic deformation, with only a small difference between the scratch depth and the final topography. In contrast, for the coated samples, the difference between the scratch and final topography curves is more pronounced, indicating elastic recovery of the surface after deformation up to a distance of approximately 300 µm, after which it stabilizes as the load increases. The results of elastic recovery show the difference between the coatings (~ 60%) and the substrate (30%), see Table 7. There are no observable discrete failures, indicating the change inside the residual scratch from coating to substrate, no delamination, or observable cracking. As the residual scratch line is visible on the coating surface of Sample 2 and Sample 4 that signifies no delamination of the coating from the substrate. From qualitative point of view, coating survived the tests performed in the given combination of load and shape of indenter. Quantitatively, one can compare the range of deformation and its elastic recovery, which is shown in Table 7. There was no visual failure.

**Table 7**. Elastic recovery during scratch test.

| Sample | Elastic recovery [%] |
|---|---|
| HVAF Run 2 | 57.5 |
| HVAF Run 4 | 61.2 |
| HVAF substrate | 30.3 |

Theoretically, superelastic NiTi alloy shall display larger elastic recovery than shape memory alloy (69-81% against 37-52% [30-32]). HVAF NiTi coating displays elastic recovery 57-61% on unloading, which is somewhere in between. This can be easily understood as being due to lattice defects and internal stress in the as-fabricated coating, which shows neither superelasticity nor shape memory. The surface prevails during these scratches as we were not able to induce cracking even at a maximum load of 500 mN, so it seems that surface coating prevailed despite the high roughness or uneven topography (as apparent from topography scratches). In general, porosity can decrease strength of coatings as it is an inducing point of cracking and failure. But this did not happen in the present scratch tests, on the contrary, cohesion of the material itself prevailed. No failure occurred due to the tip sinking into a hidden pore. Only roughness or topographical variations in height on the sample surface—likely caused by the sample's texture— which is visible around and inside the scratch, but it does not cause the coatings to fail.

#### 3.4.4 Mechanical properties evaluated by nanoindentation

Finally, as-sprayed NiTi coatings 1-4 were characterized by nanoindentation (Figs. 13-15). The recorded load-displacement responses are different from the nanoindentation responses of superelastic and shape memory NiTi coatings ( Fig. 13). They can be classified neither as superelastic nor as shape memory. They characterize the as-sprayed NiTi coatings. Since the force-indentation depth responses of coatings 1-4 are clearly different, it appears that the indentation responses depend on HVAF processing parameters.

Figures 14 and 15 provide information on the cyclic stability of indentation response (stability of the recovered depth and unrecovered depth) of NiTi coatings 1-4 at various maximum loads of 50-200 mN. It appears that the recovered depth is nearly constant (slightly decreases from cycle 1 to 10), but the unrecovered depth sharply decreases from cycle 1 to 10$^{th}$ cycle and does not evolve upon further cycling.

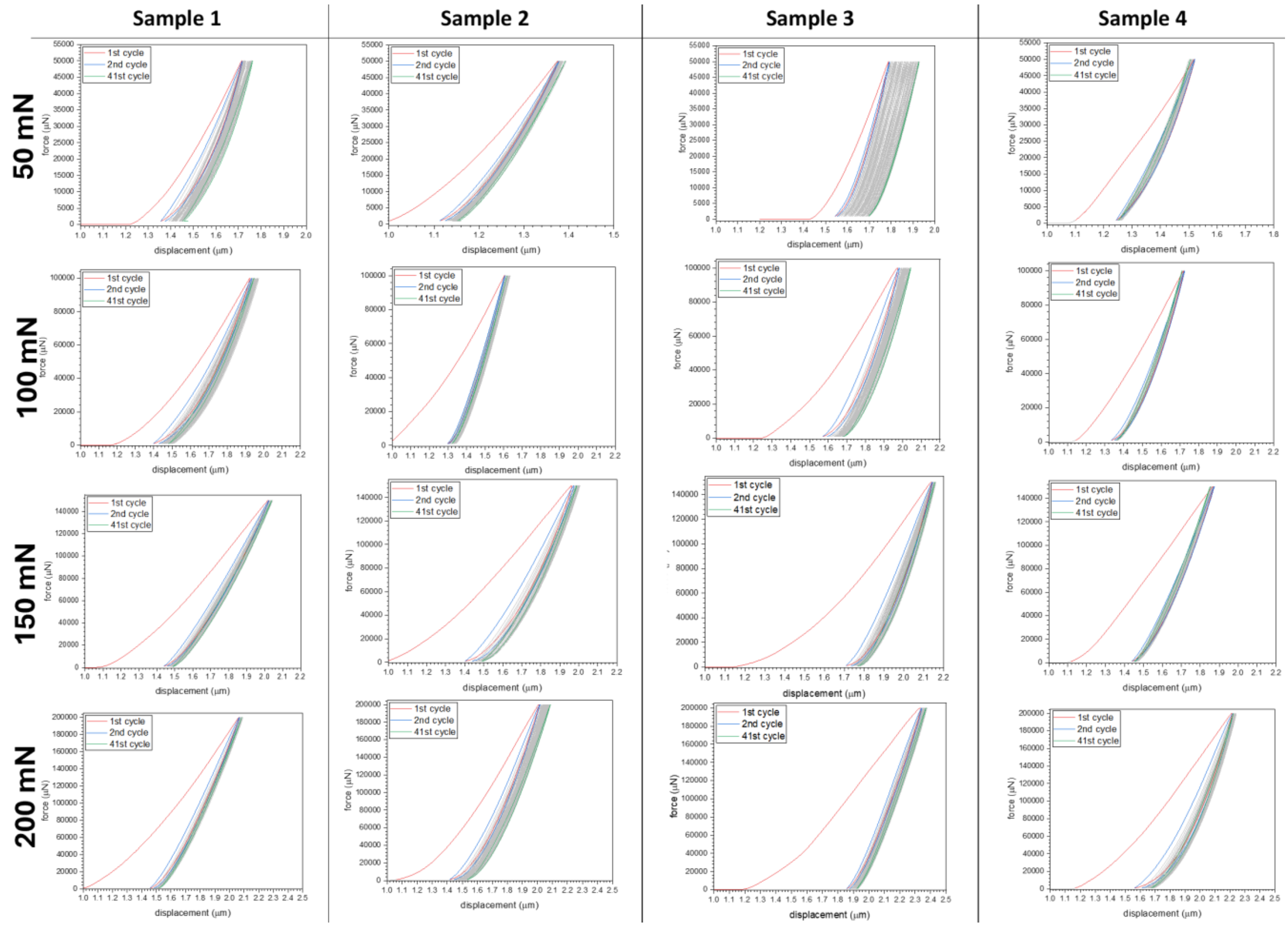

**Fig. 13:** Local mechanical response of force versus displacement evaluated in nanoindentation tests on as-sprayed HVAF coatings 1-4 performed using maximum loads 50. 100, 150, 200 mN.

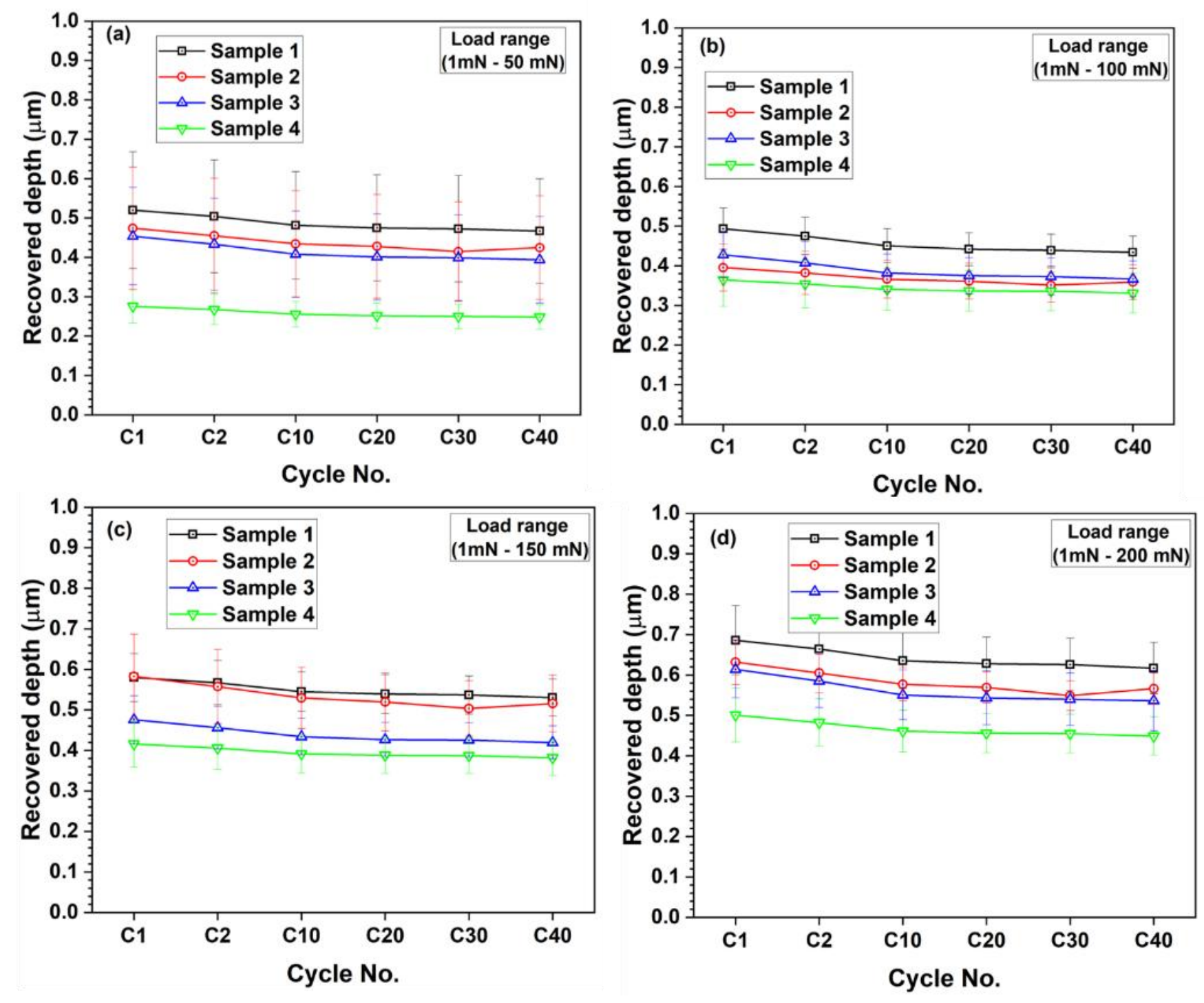

**Fig. 14.** Recovered depth evaluated from nanoindentation tests on as-sprayed HVAF coatings 1-4 at various loads in $40^{th}$ cycles. The recovered and unrecovered depth has been presented for 1, 2, 10, 20, 30 and 40 cycles. For each load (50 mN, 100 mN, 150 mN and 200 mN), the multicycle test is done at 16 different locations (in grid of 4x4 ). The average and error bars are calculated from these tests

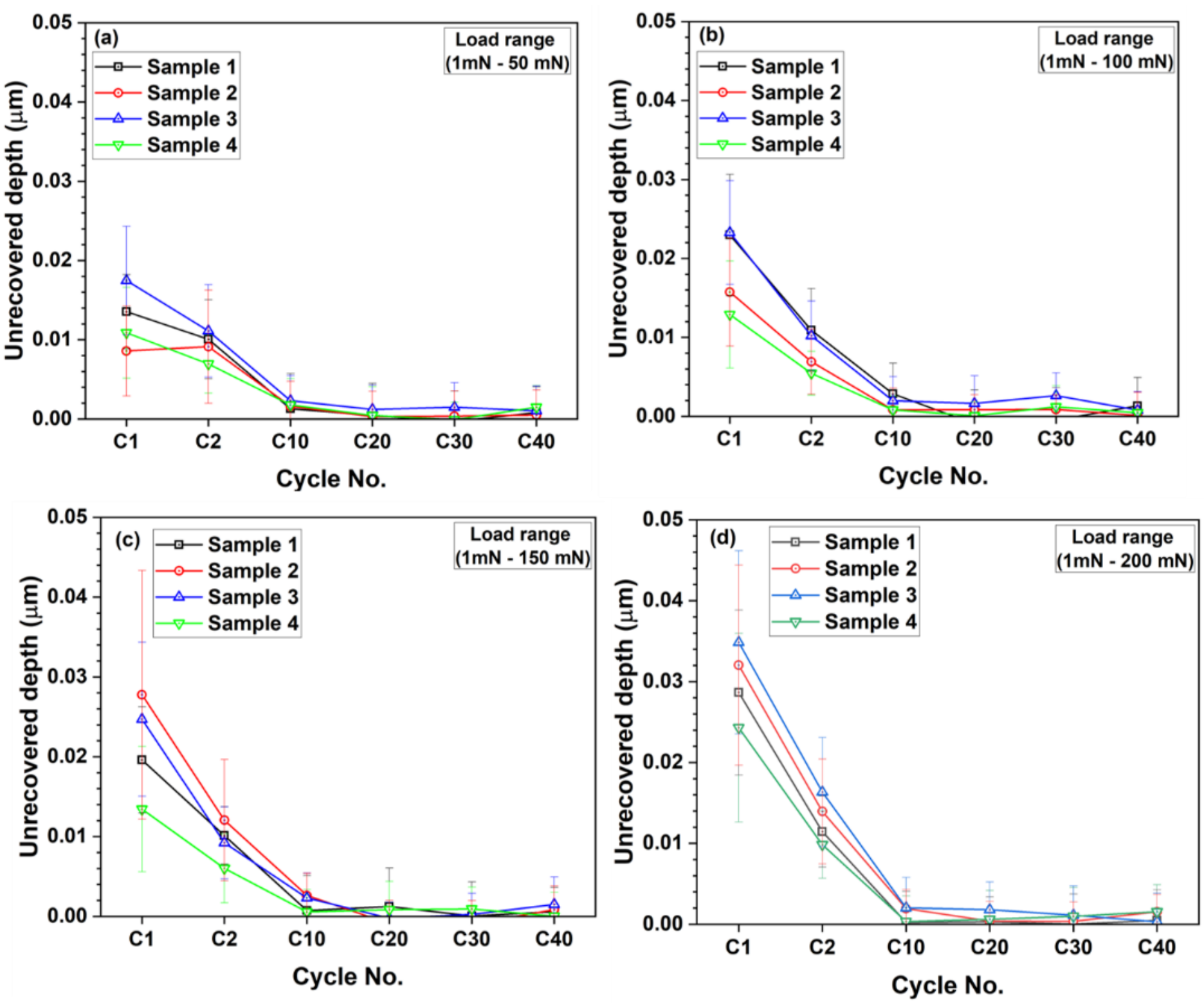

**Fig. 15:** Unrecovered depth evaluated from nanoindentation tests on as-sprayed HVAF coatings 1-4 at various loads in $40^{th}$ cycles. The recovered and unrecovered depth has been presented for 1, 2, 10, 20, 30 and 40 cycles. For each load (50 mN, 100 mN, 150 mN and 200 mN), the multicycle test is done at 16 different locations (in grid of 4x4 ). The average and error bars are calculated from these tests

## 4. Discussion

As mentioned in the introduction, deposition of thick NiTi coating onto a steel substrate creates a high-performance composite that combines the structural properties of the steel substrate with the wear resistance, corrosion protection, and shape-memory/superelastic properties of NiTi SMA. Such smart functional coatings are primarily intended for applications in medical devices, marine and heavy-duty industrial machinery, but other potentially very interesting engineering applications may appear when such NiTi coatings become readily available.

In this work, we managed to fabricate by HVAF spraying method thick NiTi coating, though the fabrication process is far from being optimized. Moreover, as-sprayed NiTi coatings need to be post processed by annealing to exhibit functional thermomechanical properties. We used annealing at 500 °C for 1 hour commonly used in the SMA field to anneal cold worked NiTi. This heat treatment changes the microstructure of the coating so that it displayes more pronounced martensitic transformations and displays functional thermomechanical properties, but again, it is not an optimized post processing treatment, possibly even better results might be achieved when applying a different heat treatment. Further research focusing on optimization of HVAF processing parameters and post processing is necessary to obtain NiTi coatings exhibiting better superelasticity and/or shape memory properties.

The results of experiments performed to characterize chemical composition, microstructure, phases, phase transformations and functional thermomechanical properties of NiTi coatings 1-4 prepared by different processing parameters were reported in section 3. We characterized the properties of four as-sprayed NiTi coatings 1-4 containing lattice defects and residual stress. If we characterized and tested systematically the annealed NiTi coatings, internal stress would be partially annealed out, but we would not know up to which extent and the results would depend on the annealing temperature and time. This was not desired, since we needed to characterize NiTi coatings in defined state.

Residual stress in HVAF NiTi coatings play an important role. They arise from two competing mechanisms that operate simultaneously while the compact layer forms during the deposition (plastic deformation, thermal contraction). The balance between them determines whether the coating is ultimately in tension or compression state which has profound consequences for adhesion, fatigue life, and the functional shape memory response.

When a semi-molten NiTi splat lands on the substrate or previously deposited layers, it deforms plastically and cools down. While cooling down, it contracts rapidly. Because the underlying material constrains this thermal contraction, tensile stresses develop within the splat. In thermal spraying hard coatings, this is the dominant mechanism, pulling the coating towards cracking and delamination. In NiTi, the situation is different since stress induced martensite forming at low temperatures may relieve these stresses. Besides of that, plastic deformation of particles can be understood as the peening effect. Since NiTi particles arrive at 600–900 m/s with kinetic energies high enough that each impact acts like a micro-hammering event on the impacted layers underneath. This plastically deforms the subsurface material and introduces compressive residual stresses, much like shot peening. As a result, HVAF NiTi coatings are expected to be under compressive-dominant stress state near the interface with the substrate, transitioning towards tension state at the free surface. This, however, needs to be evaluated experimentally. If confirmed, the gradients of internal stress might be actually favorable, since compressive stress at the interface improves adhesion strength and crack propagation resistance, while the low tensile stress at the surface is insufficient to cause spontaneous cracking. Although fabrication of NiTi coatings adherent to the substrate possessing functional thermomechanical properties is a major step forward, the quality of the fabricated coatings is far from the expected ones. Further research is underway to optimize the HVAF processing parameters and post processing heat treatments.

### 4.1 Microstructure and functional mechanical properties of the HVAF NiTi coatings

The possibility to create chemically homogeneous, single phase, nanocrystalline NiTi coating that firmly adheres to the steel substrate and undergoes B2-R-B19' MT (Figs. 3-7) is attributed to the unique characteristics of the HVAF process, which does not melt the NiTi powder particles, yet deploys sufficient amount of thermal and mechanical energies to the particles hitting the substrate under protective fuel which are sufficient to form thick NiTi layer that contains pores and oxides but transforms martensitically and displays functional thermomechanical properties (after post processing heat treatment).

The results of characterizations reported in section 3 showed that the as-sprayed coatings contain a peculiar cold work-like microstructure with nanograins, pores, oxide particles, high density of dislocations and internal stress. Major problem is the presence of pores, titanium oxides and local chemical heterogeneity in the microstructure of the as-sprayed NiTi coatings (Figs. 4,6) which make them brittle (tensile strength ~200 MPa) and prone to early fracture during thermal cycling under external stress (Figs. 10,11). These inhomogeneities were not present in the NiTi powder (except the thin natural $TiO_2$ oxide coatings on the surface of particles), they were created during the fabrication process.

HVAF uses air rather than pure oxygen as oxidizer. The peak flame temperature is roughly 1000 °C well below that of the HVOF. Less thermal energy means less thermodynamic driving force for oxidation reactions during particle heating and flight. The other advantage of HVAF is reduced dwell time. The high particle velocities in HVAF mean each particle spends far less time in the hot gas plume limiting the exposure window for oxide growth on the particle surface before impact. This is of extreme importance for highly reactive NiTi alloy which is highly prone to in-flight oxidation at elevated temperatures. Finally, there is the partial pressure effect. Even though the combustion products still contain oxygen, the lower partial pressure of oxygen in an air-fuel flame compared to an oxy-fuel systems substantially reduces the oxidation rate, which follows Arrhenius-type kinetics strongly sensitive to both temperature and oxygen activity.

Nevertheless, despite the fact that the HVAF process suppresses oxidation, the results in section 3 show presence of Titanium oxides and depletion of Ni occurring during the fabrication. Because Titanium is more reactive with oxygen than Nickel, it preferentially forms oxide particles found by TEM in the microstructure (Fig. 6). This internal oxidation consumes the Titanium from the matrix, leaving behind a proportionally Nickel-enriched matrix and intermetallic precipitates (like $Ni_3Ti$ and $Ni_4Ti_3$) in the Ni-rich matrix. At the same time, Ni depletion occurs during the thermal spray fabrication of NiTi coatings due to elemental vaporization. Because the melting point and vapor pressure of Titanium and Nickel are different, preferential vaporization of the Ni probably occurs under the influence of extreme localized temperatures during spraying which alters the final atomic stoichiometry of the deposited coating. The experimental results support this explanation. Although the NiTi coatings were found to display lower NiTi content than the NiTi powder (Table 5), the Ni content in the NiTi matrix is actually higher, since the Ti is within the Titanium oxides. Ms temperature is thus suppressed (Table 6) not just due to the plastic deformation of NiTi particles hitting the substrate, but also due to the internal oxidation occurring during the HVAF fabrication. The temperature and speed of the accelerated NiTi particles are critical process parameters.

Nevertheless, even more important is that the as-sprayed NiTi coatings do not display functional mechanical properties and have to be annealed to achieve that. The post processing annealing modifies the microstructure (Fig. 6) and changes the transformation path from the B2-B19' observed in the powder to the B2-R-B19' MT displayed by the NiTi coatings (Fig. 8). The R-phase shall not be present in Ti-rich NiTi. Its presence in as-sprayed NiTi coatings can be explained either by plastic deformation introduced during the HVAF processing and/or compositional changes due to internal oxidation.

Unfortunately, since the post processing heat treatment does not affect the porosity of the NiTi coatings, it does not increase the tensile strength. When the NiTi coating will be fabricated without pores and oxide inhomogeneities, tensile strength will increase which will open the space for exploitation of the functional thermomechanical properties. Functional properties of the coating is further improved by post processing heat treatment. Optimization of the HVAF processing (application of optimized process parameters to avoid pores and oxides) as well as post processing heat treatments is necessary to obtain NiTi coatings of desired quality. On the other hand, problems with strain compatibility at the interface between the substrate and fully annealed NiTi coatings can be expected.

We have proven that the HVAF NiTi coatings display functional mechanical properties by performing experiments on coating material separated from the substrate which is feasible due to their large thickness. This could be achieved because the HVAF fabrication enables compaction while preserving the B2 austenite phase and keeping chemical composition unchanged and macroscopically homogeneous - a challenge that has limited the scientific development of thermally sprayed smart material coatings for decades.

### 4.1 Characterization of mechanical properties of NiTi coatings by scratch and nanoindentation methods

Although the thermomechanical loading tests are excellent to prove the functional mechanical properties, we need to characterize them by surface characterization methods – i.e. without detaching the NiTi coating from the steel substrate. To achieve that, we performed scratch and nanoindentation experiments. It is found that the results (Figs. 12-15) are sensitive enough to make distinction among NiTi coatings fabricated using different processing parameters (Figs. 12,13) as well as to capture their mutually different functional fatigue performance (cyclic stability of multicycle indentation) (Figs. 14,15).

## 5. Conclusion

Thick NiTi shape memory alloy coatings (100-300 μm) were deposited on mild steel by high velocity air fuel thermal spray method. These NiTi coatings undergo B2-R-B19' martensitic transformation and display functional thermomechanical properties.

Microstructures and mechanical properties of NiTi coatings fabricated using four different sets of HVAF processing parameters were evaluated. It was found that the as-sprayed NiTi coatings display heterogeneous grain microstructures, contain voids, oxide particles, high density of dislocation defects and internal stresses that originating from the fabrication process and depend on processing parameters. The coatings are chemically homogeneous on macroscale but not on nanoscale and mesoscale, which is attributed to the internal oxidation occurring during the HVAF fabrication process.

The as-sprayed NiTi coatings need to be subjected to post processing heat treatment (we used annealing at 500 ºC for 1 hour) to display functional thermomechanical properties. Although the coatings have only limited tensile strength of ~200 MPa, the annealed NiTi coating displayed relatively large recoverable deformation in 3-point bending thermal cycling tests.

The functional thermomechanical properties of NiTi coatings on the substrates can be evaluated by surface characterization experimental methods, particularly by scratch tests and nanoindentation. Multicycle nanoindentation can be used to evaluate functional fatigue performance of the NiTi coatings. The surface characterization methods need to be further developed before they are used to optimize HVAF processing and annealing of NiTi coatings

**Declaration of Competing Interest**

The authors declare that they have no known competing financial interests or personal relationships that could have appeared to influence the work reported in this paper.

**Data Availability**

Dataset of High-Velocity-Air-Fuel spraying of NiTi coating [Data set]. Zenodo. DOI 10.5281/zenodo.21101821

**Acknowledgement**

Support from Czech Science Foundation (CSF) project 25-16285S is acknowledged. P. Šittner acknowledges support from Czech Academy of Sciences through Praemium Academiae. MEYS of the Czech Republic is acknowledged for the support of infrastructure projects, CNL (CzechNanoLab LM2023051) and FerrMion (CZ.02.01.01/00/22_008/0004591).